\documentclass[prb,twocolumn, showpacs,superscriptaddress,preprintnumbers,amsmath,amssymb]{revtex4}
\usepackage[usenames,dvipsnames]{xcolor}

\newcommand{\YSA}{Yb$_2$Si$_2$Al}

\usepackage{float}
\usepackage{graphicx}% Include figure files
\usepackage{dcolumn}% Align table columns on decimal point
\usepackage{bm}% bold math

\begin{document}

%\preprint{draft, Andrea Thomas}

\title{Intermediate valence in single crystalline Yb$_2$Si$_2$Al}

\author{W. J. Gannon}
  \affiliation{Texas A$\&$M University, Department of Physics and Astronomy, 4242 TAMU, College Station, TX 77843, USA}
\author{K. Chen}
  \affiliation{Institute of Physics II, University of Cologne, Z{\"u}lpicher Stra{\ss}e 77, D-50937 Cologne, Germany}
\author{M. Sundermann}
  \affiliation{Institute of Physics II, University of Cologne, Z{\"u}lpicher Stra{\ss}e 77, D-50937 Cologne, Germany}
  \affiliation{Max-Planck-Institute for Chemical Physics of Solids - N{\"o}thnizer Stra{\ss}e 40, 01187 Dresden, Germany}
\author{F. Strigari}
  \affiliation{Institute of Physics II, University of Cologne, Z{\"u}lpicher Stra{\ss}e 77, D-50937 Cologne, Germany}
\author{Y. Utsumi}
 \altaffiliation{Present address: Synchrotron SOLEIL, L'Orme des Merisiers, Saint-Aubin, BP 48, 91192 Gif-sur-Yvette C$\acute{e}$dex, France}
  \affiliation{Max-Planck-Institute for Chemical Physics of Solids - N{\"o}thnizer Stra{\ss}e 40, 01187 Dresden, Germany}
\author{K.-D. Tsuei}
\affiliation{National Synchrotron Radiation Research Center, 101 Hsin-Ann Road, Hsinchu 30077, Taiwan}
\author{J.-P. Rueff}
\affiliation{Synchrotron SOLEIL, L'Orme des Merisiers, Saint-Aubin, BP 48, 91192 Gif-sur-Yvette C$\acute{e}$dex, France}
\affiliation{Sorbonne Universit$\acute{e}$, CNRS, Laboratoire de Chimie Physique - Mati{\`e}re et Rayonnement, LCPMR, 75005 Paris, France}
\author{P. Bencok}
\affiliation{Diamond Light Source, Science Division, Didcot OX11 0DE, United Kingdom}
\author{A. Tanaka}
\affiliation{Department of Quantum Matter, AdSM, Hiroshima University, Higashi-Hiroshima 739-8530, Japan}
\author{A.~Severing}
  \affiliation{Institute of Physics II, University of Cologne, Z{\"u}lpicher Stra{\ss}e 77, D-50937 Cologne, Germany}
  \affiliation{Max-Planck-Institute for Chemical Physics of Solids - N{\"o}thnizer Stra{\ss}e 40, 01187 Dresden, Germany}
\author{M. C.~Aronson}
  \affiliation{Texas A$\&$M University, Department of Physics and Astronomy, 4242 TAMU, College Station, TX 77843, USA}

\date{\today}

\begin{abstract}
\YSA\ may be a prototype for exploring different aspects of the Shastry-Sutherland lattice, formed by planes of orthogonally coupled Yb ions.  Measurements of the magnetic susceptibility find incoherently fluctuating Yb$^{3+}$ moments coexisting with a weakly correlated metallic state that is confirmed by measurements of the electrical resistivity. Increasing signs of Kondo coherence are found with decreasing temperature, including an enhanced Sommerfeld coefficient and Kadowaki-Woods ratio that signal that the metallic state found at the lowest temperatures is a Fermi liquid where correlations have become significantly stronger. A pronounced peak in the electronic and magnetic specific heat indicates that the coupling of the Yb moments to the conduction electrons leads to an effective Kondo temperature that is approximately 30 K.  The valence of Yb$_2$Si$_2$Al has been investigated with electron spectroscopy methods. Yb$_2$Si$_2$Al is found to be strongly intermediate valent ($v_F$\,=\,2.68(2) at 80\,K). Taken together, these experimental data are consistent with a scenario where a coherent Kondo lattice forms in \YSA\ from an incoherently fluctuating ensemble of Yb moments with incomplete Kondo compensation, and strong intermediate valence character.
\end{abstract}

 %wjg 2/28/2018

\maketitle

\section{Introduction}

Quantum magnetism is one of the most dynamic fields in condensed matter physics, and there is a particular focus on finding and exploring ground states where quantum fluctuations are so strong that they overcome magnetic order.\cite{Powell2011,Lee2014,Savary2017,Zhou2017} The ground state of these systems is a quantum mechanical superposition of local states, such as singlet pairs of local moments, leading to a ground state wave function that is massively entangled. Unlike a conventional phase transition, which is accompanied by a broken symmetry, these systems instead host topological phase transitions where there is a discontinuity in a global topological invariant, such as the Chern number of the energy bands, or the flux of an emergent field. Much interest has focused on the novel excitations of these systems, as they can have fractional or anionic statistics with pronounced topological properties.

So far, most of what we know about these systems comes from experimental and theoretical results for insulators where magnetic moments lie on a geometrically frustrated lattice. Following Anderson's initial proposal\,\cite{Anderson1973,Anderson1987,Lee2006} that a  frustrated and non-ordered magnetic state, the ``Resonant Valence Bond'' state, could be doped with mobile carriers to produce high temperature superconductivity, there has been great interest in discovering metallic systems without long ranged  magnetic order. The most direct approach is to introduce charged impurities into the insulating host,\cite{Mazin2014,Lee2006} however, the associated carriers are almost always localized.\cite{Dabkowska2007,Kelly2016} An alternative approach is to seek intrinsically metallic systems where magnetic moments lie on a geometrically frustrated lattice. Perhaps the earliest report of such a system was the allotrope $\beta$-Mn, which has a pyrochlore structure~\cite{Stewart2002,Paddison2013}, and there is a  more recent report of Sc$_{3}$Mn$_{3}$Al$_{7}$Si$_{5}$, where Mn moments lie on a Kagome lattice.\cite{He2014} There is  an increasing number of systems where rare earth moments lie on geometrically frustrated lattices, such as pyrochlore Pr$_{2}$Ir$_{2}$O$_{7}$,\cite{Nakatsuji2006} and the Kagome lattice systems CePdAl\,\cite{Fritsch2017,Lucas2017} and YbAgGe.\cite{Tokiwa2013}

We will focus here on a different frustrated lattice, where rare earth moments form planes of orthogonal dimers in the Shastry-Sutherland Lattice (SSL).\cite{Shastry1981} One of the earliest systems where an exact solution was identified, it is the interplay between the stability of local spin singlets, enabled by the nearest-neighbor interaction $J$, and the tendency towards long ranged order that is driven by interdimer coupling $J^{\prime}$ that leads to a complex array of ordered and disordered states in the SSL plane.\cite{Miyahara2003} The first SSL systems that were extensively explored were the metallic  $R\mathrm{B}_{4}$ ($R$\,=\,Gd,Tb,Dy,Ho,Er) materials, however quantum fluctuations are apparently very weak in these compounds, all having  conventional antiferromagnetic ground states.\cite{Etourneau1979,Fisk1981,Michimura2006,Iga2007,Matas2010} The discovery\,\cite{Kim2008,Kim2011,Kim2013} that SSL physics is present in the $R_{2}T_{2}X$ compounds was a significant step forward, as this is an extensive family\,\cite{Lukachuk2003,Rodewald2007} with many compositional variations that depend on the rare earth $R$ and transition metal $T$.  Depending on the ratio of the distances between dimers in the SSL plane and between SSL planes, the $R_{2}T_{2}X$ compounds can display either one or two dimensional magnetic behaviors.\cite{Kim2008,Kim2011} Remarkably, the observation of fractionalized excitations in  Yb$_{2}$Pt$_{2}$Pb\,\cite{Wu2016} is firm evidence that this is a one-dimensional system where interchain coupling is  minimized by the orthogonality of the Yb dimer wave functions, a last remnant of SSL physics.

There is considerable theoretical evidence that unconventional superconductivity is stable through much of the SSL phase diagram, especially near the quantum critical point (QCP) $(J/J^{\prime})_{QCP}$\,=\,1.43 where magnetic order vanishes. Electronic correlations play a crucial role in the nature of the superconducting pairing.  If the near-neighbor Coulomb repulsion is weak, the mobile carriers will be localized on individual dimers, leading to $s$-wave superconductivity.\cite{Yang2008}. If repulsion is strong, then only one carrier resides on a dimer, and it is the long-ranged correlations among dimers that leads to $d$-wave superconductivity.\cite{Shastry2002,Chung2004} Computations of the phase diagram for rare earth moments on the SSL\,\cite{Bernhard2011,Pixley2014,Su2015} have emphasized the important role of the hybridization $\cal{J}$ between the conduction electrons and $f$-electron moments in enabling unconventional superconductivity on the SSL.

The central question is whether the novel excitations that are possible in one and two dimensional systems based on the SSL persist as the hybridization $\cal{J}$ increases, perhaps leading to unconventional superconductivity, or if it is the quenching of the rare earth moments that enables superconductivity, as has been demonstrated in a number of heavy fermion compounds. It is necessary to identify a family of compounds where it is possible to explore the impact of increasing $\cal{J}$ in a structural motif that has already been proven to host fractionalized excitations.

We have selected the compound Yb$_{2}$Si$_{2}$Al for this study, since the hierarchy of length scales\,\cite{Kranenberg2000} suggests that the intradimer interaction in the SSL plane would be largest (near-neighbor spacing =\,3.39\,\AA), followed by the interdimer interaction (spacing =\,3.52\,\AA) in the plane, with the interdimer interaction between planes the weakest (spacing = 4.23\,\AA.) This implies that the magnetism in  \YSA\ may be quasi-two dimensional, and perhaps dominated by SSL physics.  As desired, the reported magnetic properties of \YSA\ suggest that the Yb $f$-electron and conduction electron hybridization $\cal{J}$ is stronger than in Yb$_2$Pt$_2$Pb and that Yb is in an intermediate valence state,\cite{Shah2009} i.e. the ground state of each Yb ion is a quantum mixture of the Yb$^{3+}$ ($f^{13}$) and Yb$^{2+}$ ($f^{14}$) configurations.  While this previous work was carried out on polycrystalline samples, we have been able to synthesize high-quality single crystals. We find a modest magnetic anisotropy, most likely due to the single ion anisotropy of the Yb ions themselves. Electron spectroscopy and x-ray absorption measurements were  used to investigate the valence and ground state symmetry of the Yb ions, confirming that the Yb ions in \YSA\ are strongly intermediate valent. The temperature dependencies of the magnetic susceptibility and specific heat indicate that \YSA\ displays Kondo lattice behavior, with incoherently fluctuating Yb moments at high temperature, where the onset of Kondo coherence among these moments results in a heavy Fermi liquid ground state that is evidenced in the large Sommerfeld coefficient and the enhanced Kadowaki-Woods ratio. In these respects, \YSA\ is very similar to other Yb-lattice systems where the Yb ions are in an intermediate valence state.

\section{Sample preparation}

Single crystals of \YSA\ in a rodlike morphology were grown from an Al self flux. X-ray diffraction measurements were carried out on powder of \YSA\ produced from these single crystals, confirming the tetragonal Mo$_2$FeB$_2$ structure type that was previously reported.\cite{Kranenberg2000, Shah2009} Single crystal x-ray diffraction experiments showed that the (001) crystal axis is oriented parallel to the long axis of the rod, perpendicular to facets that define the (110) direction.  Typical crystal dimensions are $\approx$\,0.5\,mm along (110) and several millimeters along (001).  Wavelength dispersive spectroscopy performed on single crystals confirmed the nominal Yb$_2$Si$_2$Al composition, which showed no variation across the crystal surface within the $\simeq$\,1\%  experimental uncertainty.

\section{Macroscopic characterization}

\subsection{Magnetization \& Susceptibility}
\begin{figure}[t]
    \centering
    \includegraphics[width=0.9\columnwidth]{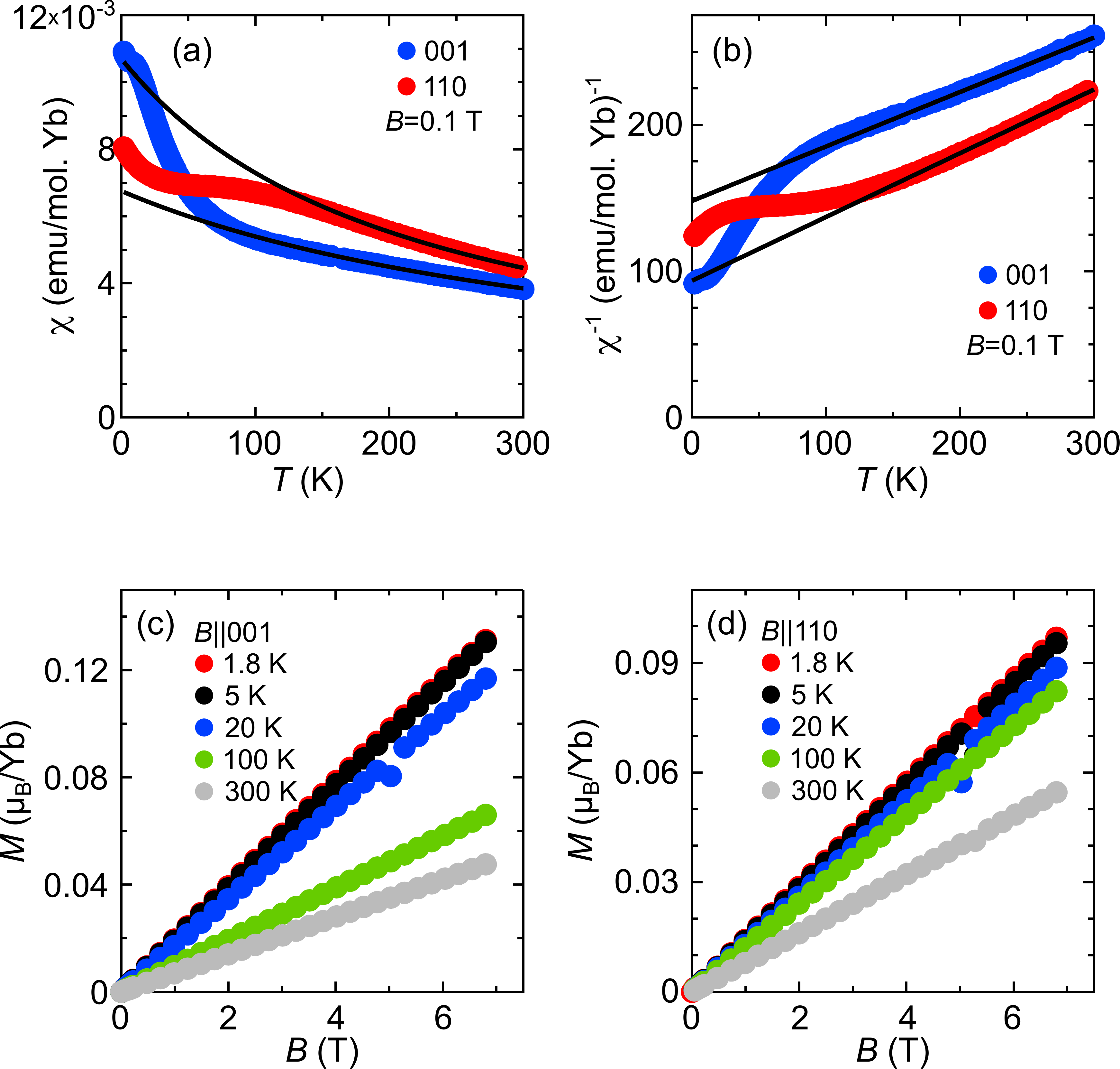}
		\caption{(a) The temperature dependence of the DC magnetic susceptibility $\chi(T)=M/B$, where the magnetization $M$ is measured in a fixed field $B$\,=\,0.1\,T that is oriented along the (001) (blue) and (110) (red)  directions.  Black lines are fits of the inverse susceptibility $\chi^{-1}$ to the Curie-Weiss law. (b) The temperature dependencies of  $\chi^{-1}$ measured with a fixed field $B=0.1$ T along the (001) (blue) and (110) (red) directions. Black lines are fits of $\chi^{-1}$ to the Curie-Weiss law for $T\,>\,150$ K.  (c and d) The isothermal magnetization $M(H)$ is measured as functions of the magnetic fields $H$ oriented along the (001) direction (c) and (110) direction (d) at different fixed temperatures $T\,=\,1.8$ (red), 5 (black), 20 (blue), 100 (green), and 300\,K (gray).}
    \label{fig1}
\end{figure}

Measurements of the magnetization M were obtained for temperatures $1.8~\mathrm{K}\leq~T~\leq~300~\mathrm{K}$ using a Quantum Designs Magnetic Property Measurement System (MPMS).  The temperature dependence of the DC magnetic susceptibility $\chi(T)=M/B$ was measured in a fixed field of $B=0.1~\rm T$ (Fig.~\ref{fig1}\,(a)). The susceptibility $\chi$ varies by no more than 4\% at any temperature when the 0.1~T field is oriented within the \textit{ab}-plane. However, there is a pronounced and temperature dependent axial anisotropy in \YSA\, since $\chi$ is smaller when the field is oriented along the (001) $c$-axis than along the (110) direction in the basal plane at high temperatures, and this anisotropy is reversed for temperatures $T<45~\rm K$. Overall, the temperature dependence of the susceptibility $\chi(T)$ is relatively weak, although a strong upturn at low temperatures is evident, particularly for fields perpendicular to the basal plane.  There is no overt evidence in $\chi(T)$ or its temperature derivative that would indicate magnetic order above 1.8~K.  Apart from the information about the magnetic anisotropy, the overall temperature dependence and magnitude of the susceptibility data are qualitatively consistent with previous reported measurements that were carried out on a powdered sample.\cite{Shah2009}

Above $\simeq$100\,K, the magnetic susceptibility in \YSA\ is well fitted by the Curie-Weiss temperature dependence, $\chi(T)=C/(T+ \theta_{CW})$ (Fig.~\ref{fig1}(a,b)). The Weiss temperatures $\theta_{CW}$ extracted from these fits are large and negative for fields in the basal plane and along the $c$-axis, with $\theta_{CW}$\,=\,-397\,$\pm$\,5\,K for $B\|(001)$ and $-214 \pm 4~\rm K$ for $B\|(110)$.  In an insulating magnet, this would suggest the presence of substantial and anisotropic antiferromagnetic interactions  between the Yb magnetic moments, and typically $\theta_{CW}$ would be very small, due to the small de Gennes factor for Yb.  Thus, it seems more likely that $\theta_{CW}$ in \YSA\ instead represents  the effective exchange coupling between localized Yb moments and itinerant conduction electrons. The effective moments extracted from the Curie constant $C$ are 4.27\,$\pm$\,0.02\,$\mu_B$/$\mathrm{Yb}$ for $B\|$(110) and 4.63\,$\pm$\,0.02\,$\mu_B$/$\mathrm{Yb}$ for $B\|$(001), comparable to the  4.54\,$\mu_B$/$\mathrm{Yb}$ that Hund's rules give for the ground state of Yb$^{3+}$.

Below about $\simeq 150~\rm K$ ($B\|$(110)) and $\simeq 80~\rm K$ ($B\|$(001)) the temperature dependent part of the susceptibility no longer follows the Curie-Weiss law, with this deviation having a broad maximum in the basal plane susceptibility at $\simeq 100~\rm K$ for $B\|$(110) (Fig.~\ref{fig1}(a)). A strong upturn in the $c$-axis susceptibility, and a more moderate upturn in the basal plane susceptibility are the dominant features below 100~K. In principle, this upturn could be related to the presence of trace amounts of Yb$_{2}$O$_{3}$ in the samples, but the anisotropy in the susceptibility upturn in \YSA\ (Fig.~\ref{fig1}~(a)) suggests this is not a significant effect in our single crystal samples. As we will argue below, it is more likely that the anisotropy corresponds largely to the single ion anisotropy, reflecting the nature of the doublet ground state and the occupation of the crystal-electric field manifold for the Yb ions.  The Curie-Weiss fits to the susceptibility are not improved by the addition of a temperature independent offset, indicating that the Pauli susceptibility is small, as befits a simple metal with weak electronic correlations.

The isothermal magnetization curves  $M(H)$ (Fig.\,\ref{fig1}\,(c,\,d)) are linear at all temperatures from 1.8\,-\,300\,K, and for fields as large as 6.8\,T. Since there is no evidence in \YSA\ for magnetic order at any temperature below 300\,K, the magnetization isotherms suggest instead that the magnetic moments are strongly hybridized with the conduction electrons, so that nonlinear field dependencies would only become evident when the Zeeman splitting approaches the characteristic energy scale for the hybridization.\cite{Hewson1985}

The breakdown of the Curie-Weiss temperature dependence found at high temperatures, the broad maximum in $\chi(T)$, and the linear field dependence of the magnetization, are all features that are typical of metallic systems where Yb has intermediate valent character.\cite{Sales1975,Klaasse1981} Near and above room temperature, the Yb moments in \YSA\ are significantly coupled to the conduction electrons but in an incoherent manner, analagous to Kondo impurities. With decreasing temperature, coherence among the fluctuating Yb moments develops and \YSA\ evolves towards a Fermi liquid ground state, where the Yb moments are substantially less magnetic and the Fermi liquid becomes strongly correlated.\cite{Fisk1995,Khomskii}

\begin{figure}[t]
    \centering
    \includegraphics[width=0.8\columnwidth]{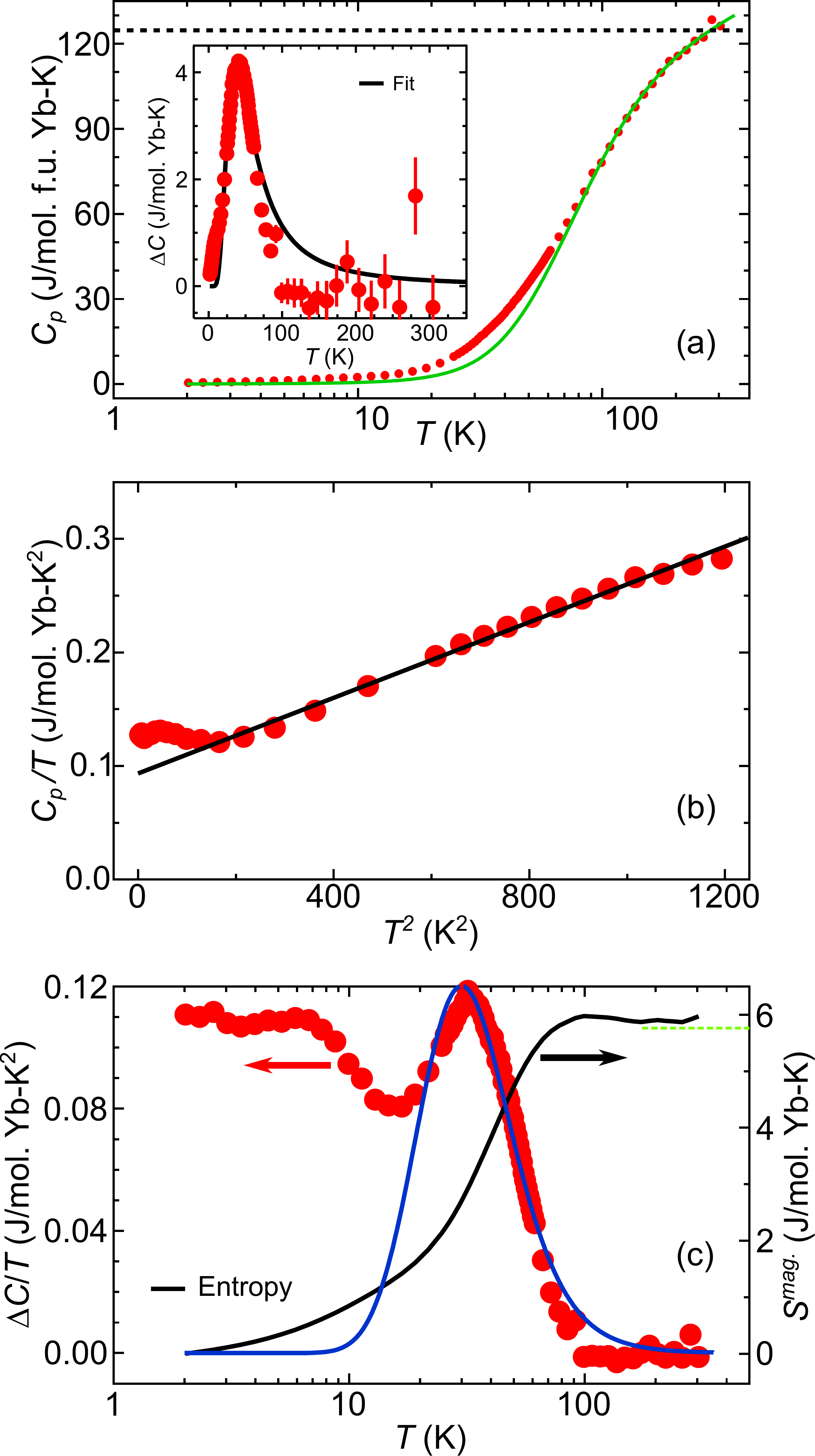}
		\caption{(a) The total specific heat per \YSA\ formula unit $C_p$ is measured as a function of temperatures between  1.8\,$<$\,$T$\,$<$\,300\,K (red circles).  A fit to the Debye model with an electronic contribution $\gamma_{HT}T$  (see text) is shown (green line) along with the Dulong-Petit limit of $3R$ per mole of each atom, where $R$ is the ideal gas constant (black dashed line).  (a, inset) The excess specific heat per Yb ion $\Delta C$\,=\,$C_{p}$\,-\,$C_{ph}$ divided by temperature $\Delta C/T$ as a function of temperature.  The peak in $\Delta C/T$ is fitted to a Schottky expression (black line, see text). (b) The total specific heat divided by temperature $C_p/T$ as a function of temperature squared, $T^2$.  A fit to the function $C_p/T$\,=\,$ \gamma$\,+,$\beta T^2$ over the range 16\,$\leq$\,$T$\,$\leq$\,35\,K is also shown (black line). (c) The excess specific heat per Yb ion divided by temperature $\Delta C/T$ as a function of temperature (red circles, left axis) and the excess entropy $S^{\mathrm{mag}}$ per Yb ion as a function of temperature (black line, right axis) (see text.)  $R\mathrm{ln}2$ of entropy is marked by the green dashed line on the right axis.  The Schottky  fit from part a, inset is also shown (blue line).}
    \label{fig2}
\end{figure}

\subsection{Specific Heat}
The temperature dependence of the zero field specific heat, $C_p$ was measured in a Quantum Designs Physical Property Measurement System (PPMS) (Fig.~\ref{fig2}). In order to isolate the magnetic and electronic contributions to the specific heat, a Debye model is used to estimate the contribution to $C_p$ from phonons, $C_{ph}$. As shown in Fig.\,\ref{fig2}\,(a), $C_{p}$ is well fitted for $T$\,$>$\,100\,K by selecting a Debye temperature $\Theta_D$\,=\,332\,$\pm$\,3\,K, and a small electronic contribution $\gamma_{HT}T$, where $\gamma_{HT}\,=\,0.016 \pm 0.001~\mathrm{J/mol\,Yb\cdot K^2}$. The small values of $\gamma_{HT}$ and inferred Pauli susceptibility indicate that the metallic background present at high temperatures is a weakly correlated electronic fluid.

The electronic and magnetic specific heat $\Delta$$C$\,=\,$C_p$\,-\,$C_{ph}$ is presented in Fig.\,\ref{fig2}\,(a, inset).  $\Delta$$C$ vanishes above $\simeq$\,100\,K, but with reduced temperature there is a distinctive peak centered at $T\simeq 40~\rm K$, indicating the presence of a characteristic energy scale, which could refer either to the Kondo temperature of the Yb ions, or to the coherence temperature of the lattice. The peak is fitted to a Schottky expression, which gives a relative degeneracy of three for the two states, which are separated by a temperature scale of (107\,$\pm$2)\,K.  As $T\rightarrow0$, the plot of $C_{p}/T$ as a function of $T^{2}$ (Fig.~2b) shows that the Sommerfeld coefficient of the $T\rightarrow0$  electronic specific heat $\gamma$\,=\,0.091\,$\pm$\,0.003\,J/mol\,Yb$\cdot$K$^2$ has increased by approximately a factor of six relative to the value $\gamma_{HT}$ found at high temperature.  Plotting $\Delta C/T (T)$ (Fig.\,\ref{fig2}(c)) shows that $\Delta C/T$ ultimately saturates for $T$\,$<$\,8\,K, allowing a separate determination of the Sommerfeld constant $\gamma$\,=\,0.110\,J/mol\,Yb$\cdot$K$^2$.  Both methods for determining $\gamma$ are in reasonable agreement, indicating that the isolation of $\Delta$$C$ is not overly influenced by the procedures used to estimate $C_{ph}$.  The specific heat data show that a strongly correlated Fermi liquid develops at the lowest temperatures in \YSA\, with a Sommerfeld constant $\gamma$ that has a magnitude that is typical of other Yb Kondo lattice compounds.\cite{Fisk1995,Bleckwedel1981,Tsujii2005,Nakatsuji2008,Matsumoto2011,Fernandez2012,Jiang2015}

The magnetic susceptibility and specific heat measurements are overall consistent with \YSA\ having a substantial degree of hybridization $\cal{J}$ between the Yb moments and the conduction electrons. At the highest temperatures, Yb$^{3+}$  atoms with the full Hund's rule moment exhibit single ion behavior within a metallic background with minimal electronic correlations. Below a characteristic temperature of $\simeq$\,15 K, these moments have been co-opted into a collective state, where their interaction with the conduction electrons leads to a strongly interacting electronic fluid that seems to coexist with more localized states with excitations that suggest an origin in the quenching of individual Yb moments within the Kondo lattice. These observations are consistent with those found in Yb-based intermediate valence systems,\cite{Fisk1995,Bleckwedel1981,Tsujii2005,Nakatsuji2008,Matsumoto2011,Fernandez2012,Jiang2015} suggesting that the hybridization $\cal{J}$ is considerably stronger in \YSA\ than in nearly isostructural Yb$_2$Pt$_2$Pb. The subsequent electron spectroscopy and x-ray absorption measurements will give evidence for the strong intermediate valent character of Yb in \YSA.

\begin{figure}[t]
    \centering
    \includegraphics[width=0.8\columnwidth]{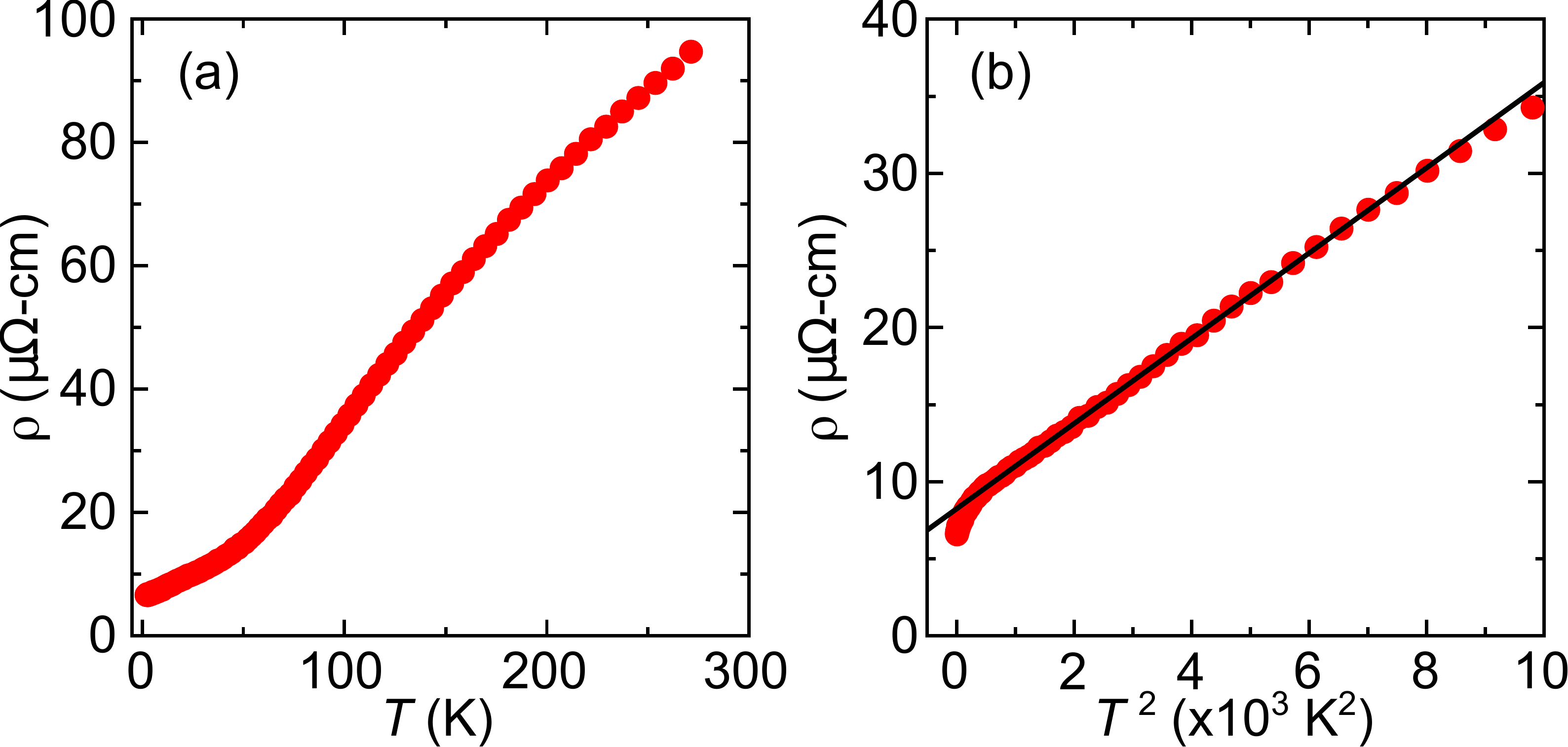}
		\caption{(a) The electrical resistivity $\rho$ as a function of temperature of \YSA, $\rho\left(T\right)$. (b) The electrical resistivity as a function of temperature squared, $T^2$ is compared to a fit of $\rho(T)$ in the low temperature region (10 K$\leq T \leq$150 K) to the Fermi liquid expression $\rho$($T$)\,=\,$\rho_{0}$\,+\,$AT^{2}$(black line).}
    \label{fig3}
\end{figure}

\subsection{Electrical Resistivity}

The temperature dependence of the electrical resistivity $\rho$($T$) was measured in a PPMS at temperatures between 300 and 2\,K, with a 500\,$\mu$A current parallel to (001)( Fig.\,\ref{fig3}a).  $\rho\left(T\right)$ shows metallic behavior at all temperatures, with the typical bulge between 100\,K\,-\,200\,K, suggestive of spin fluctuations.  At lower temperatures, the resistivity follows a quadratic temperature dependence $\rho(T)=\rho_{0}+AT^{2}$ (Fig.\,3b). The small value of $\rho_0$\,=\,8.38\,$\pm$\,0.04\,$\mu\Omega\cdot\mathrm{cm}$ confirms the quality of our single crystal samples, while the relatively large value of the coefficient $A$\,=\,0.00271\,$\pm$\,10$^{-5}$\,$\mu\Omega\cdot\mathrm{cm}\cdot\mathrm{K}^{-2}$ is consistent with a correlated ground state. Combined with the specific heat analysis, the resistivity results imply a Kadowaki-Woods (KW) ratio~\cite{Kadowaki1986} $A/\gamma^2$\,=\,0.31\,$\mu\Omega\cdot$cm$\cdot$(Yb\,mol$\cdot$K/J)$^2$. This value of the Kadowaki-Woods ratio is completely consistent with those found in other Yb compounds, but is  more than an order of magnitude lower than the values found in many heavy fermion compounds~\cite{tsujii2003, Tsujii2005}. It has been noted that this difference can be traced to the generally larger degeneracy of the Yb moments, relative to those found in heavy fermion Ce compounds.

\section{Yb Valence}
The intermediate valent character of Yb$_2$Si$_2$Al has been established with hard x-ray (h$\nu$$_{in}$\,=\,6.47\,keV) valence band and 3$d$ core level photo electron spectroscopy (HAXPES) and the temperature dependence has been measured in detail with L-edge x-ray absorption in the high resolution partial fluorescence yield (PFY-XAS) and resonant x-ray emission (RXES) (h$\nu$$_{in}$\,$\approx$\,8.5\,keV). Both methods are bulk sensitive due to the hard x-rays. The methods are based on the presence of a core hole that acts differently on the two valence configurations that form the intermediate valent ground state ($c_{13}f^{13}$\,+\,$c_{14}f^{14}$). Accordingly, spectral features ($\underline{c}f^{13}$ and $\underline{c}f^{14}\underline{L}$ with $\underline{c}$ representing a core hole) appear in the XAS or PES spectra. The intensities $I(\underline{c}f^{13})$ and $I(\underline{c}f^{14}\underline{L})$ (referred to as $I(f^{13})$ and $I(f^{14})$ for simplicity) of these structures can be related to the weights ($c_{13}^2$, $c_{14}^2$) of the $f$ states in the initial state. For Yb\,\cite{Kummer2011} final state effects are much smaller than, for example Ce so that we omit an Anderson impurity model in the formalism\,\cite{Anderson1961} of Gunnarson and Sch\"onhammer.\cite{Gunnarsson1983} Instead we assume that the $f^{13}$ and $f^{14}$ related spectral weights resemble the respective proportions in the ground state with $I(f^{13})$/[$I(f^{13})$\,+\,$I(f^{14})$]\,$\approx$\,$c_{13}^2$\,=\,$\underline{n}_f$ with valence $v_f$\,=\,2\,+\,$\underline{n}_f$.

\begin{figure}[H]
    \centering
    \includegraphics[width=0.9\columnwidth]{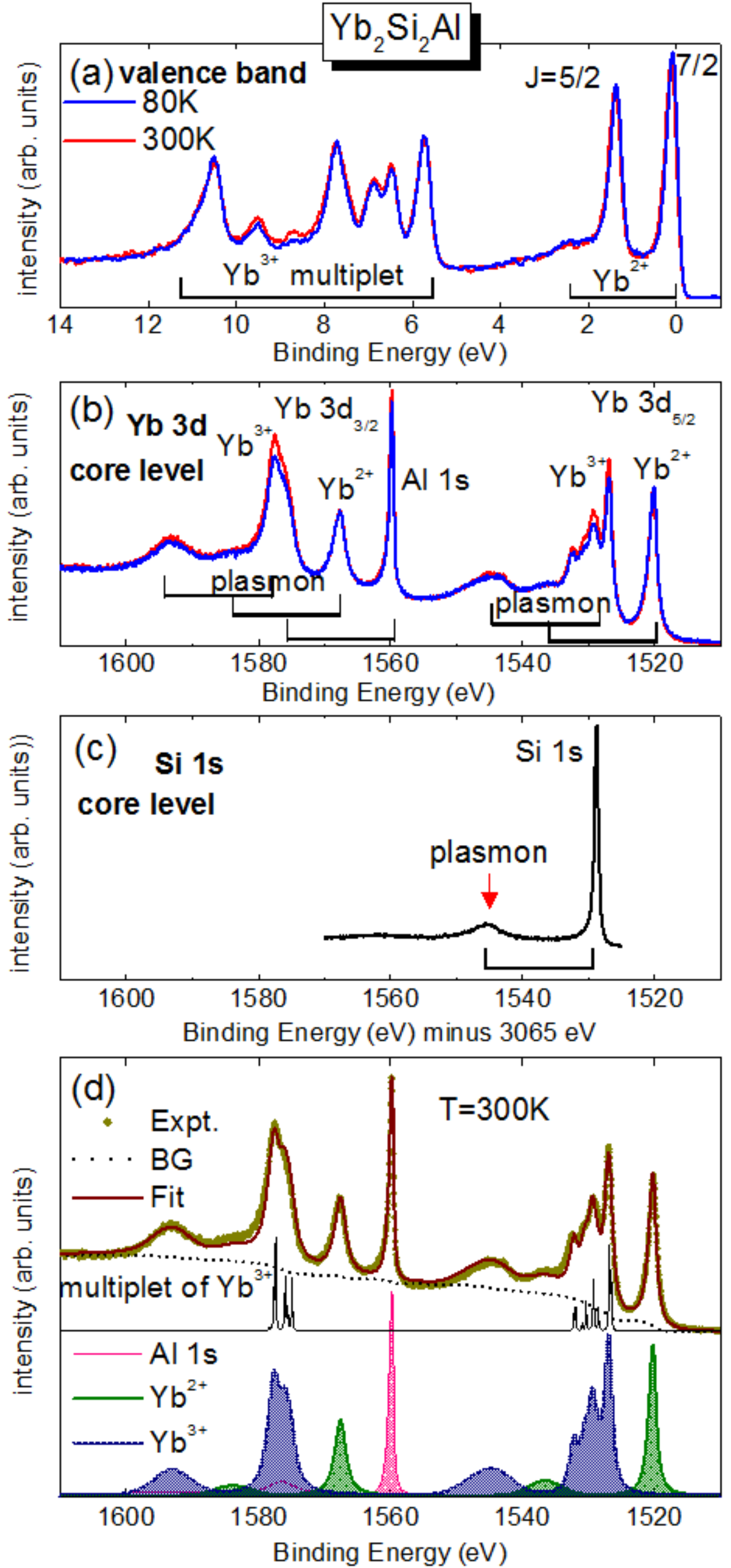}
\caption{HAXPES data of Yb$_2$Si$_2$Al measured at 80 and 300\,K with h$\nu$\,=\,6.47\,keV incident photons and a resolution of 250\,meV: (a) valence-band, (b) Yb\,3$d$ core level plus Al\,1$s$, and (c) Si 1$s$ core level. Note, the scale of the Si binding energy has been shifted by 3065\,eV. The red arrow markes the plasmon position at 16\,eV higher binding energy. (d) Assignment of spectral weights after background correction (black dotted line). The spectral shape of Yb$^{3+}$ has been reproduced with a full a multiplet calculation taking into account the plasmon intensities at 16\,eV higher binding energies.}
    \label{fig4}
\end{figure}

\subsection{HAXPES: absolute values}
The beamline and experimental set-up for the HAXPES experiment are described in the \textsl{Appendix}. Data of the valence band (Fig.\,\ref{fig4}\,(a)) and 3$d$ core levels (Fig.\,\ref{fig4}\,(b)) were taken at 80 (blue) and 300\,K (red). The respective configurations and multiplet structures are marked in the figures.

%core level is better than valence band : \cite{Sato2004, Yamaguchi2009} }
In the valence band data in Fig.\,\ref{fig4}\,(a) the multiplet structures of both Yb configurations are visible, the Yb$^{2+}$ 4$f_{7/2}$ and 4$f_{5/2}$ peaks near the Fermi energy $E_F$ and at 1.5\,eV, and the Yb$^{3+}$ 4$f$ multiplet structure in the energy range of 5 to 12\,eV. Hence the material has a valence between two and three. There is a small but distinct transfer from Yb$^{2+}$ to Yb$^{3+}$ with rising temperature (compare red and blue lines).

Fig.\,\ref{fig4}\,(b) shows the Yb\,3$d_{3/2}$ and 3$d_{5/2}$ core level spectra and also here both Yb configurations are present; single lines at 1520 and 1567\,eV for Yb$^{2+}$ and broad multiplet structures that arise from Coulomb interaction between the 3$d$ and 4$f$ holes in the electron configuration of the 3$d^9$4$f^{13}$ final states. Also here we find more Yb$^{3+}$ intensity at 300\,K with respect to 80\,K (compare red and blue lines). The pronounced humps at 1545 and 1592\,eV are attributed to plasmon intensities and have to be assigned to all emission lines for a quantitative analysis.\cite{Strigari2015, Sundermann2016} The plasmon position and relative intensity with respect to the main emission lines can be determined from the Si\,1$s$ single emission line in Fig.\,\ref{fig4}\,(c) that was measured at the same sample.

The core level and valence band data resemble the spectra of other intermediate valent Yb compounds.\,\cite{Sato2004,Suga2005,Yamaguchi2009,Suga2009,Okawa2010,Utsumi2012} For a quantitative analysis of the present core level spectra we have simulated the spectral shapes with a full multiplet calculation using the XTLS 9.0 program\,\cite{Tanaka1994} after subtracting an integrated background\,\cite{Shirley1972} (black dotted curve) and broadening in due consideration of the  plasmon contributions.\cite{Strigari2015, Sundermann2016} The resulting spectral weights yield an Yb valence of $v_f$\,=\,2.68(2) and 2.71(2) for $T$\,=\,80 and 300\,K, respectively.

\begin{figure}[t]
    \centering
    \includegraphics[width=0.9\columnwidth]{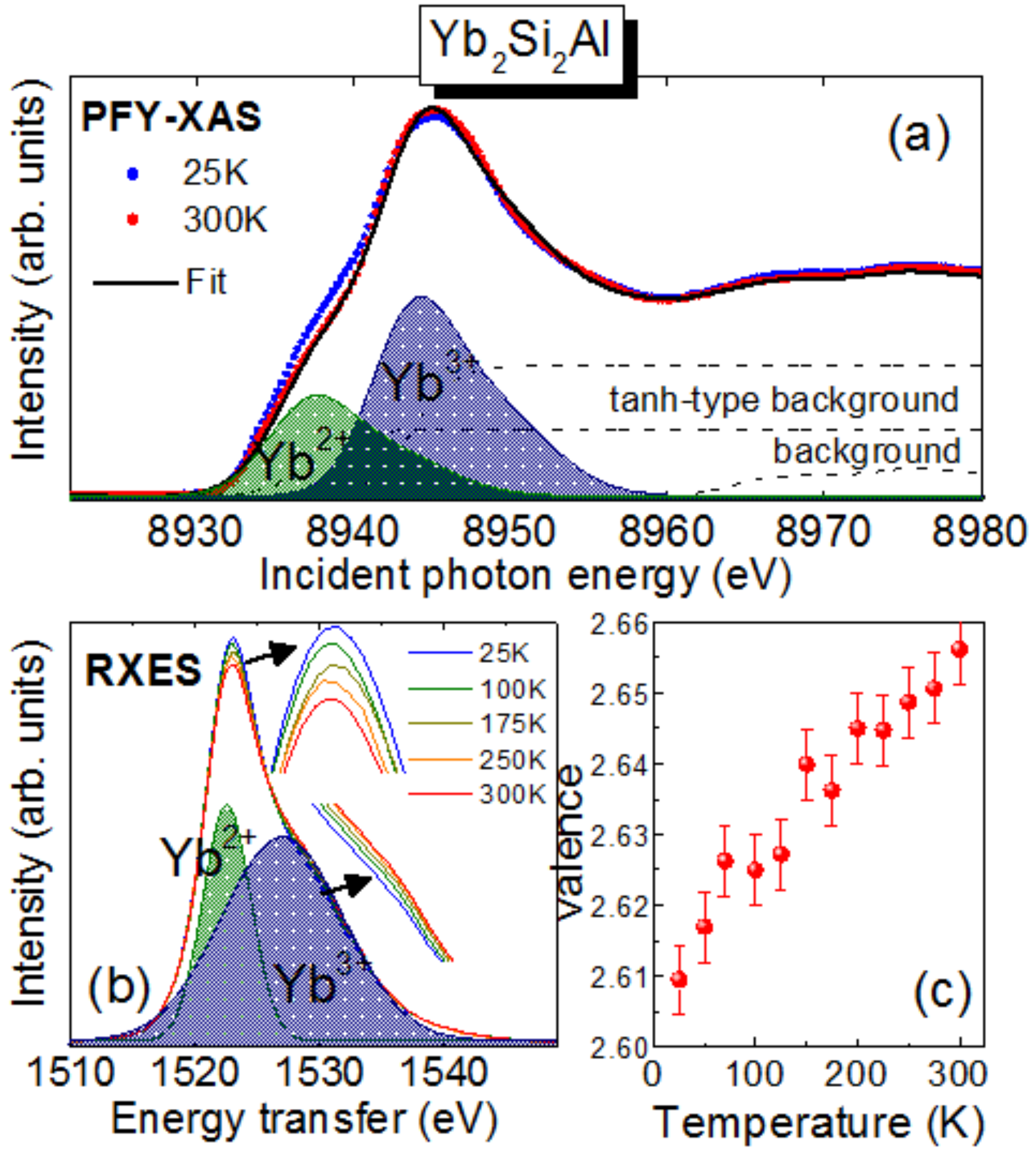}
    \caption{(a) PFY-XAS L$_{\alpha1}$ data of Yb$_2$Si$_2$Al for T\,=\,25 (blue dots) and 300\,K (red dots), and fit to 300\,K data (black line) (b) RXES data for at $h\nu$= 8938\,eV  incident energy for several temperatures and empirical assignment of spectral weights. Insets: regions of Yb$^{2+}$ and Yb$^{3+}$ peak positions on extended scales. (c) valence as function of temperature.}
    \label{fig5}
\end{figure}

\subsection{High resolution L-edge XAS: temperature dependence}
The temperature dependence has been studied in greater detail at the Yb L-edge with the PFY-XAS and RXES. In the PFY mode the incident energy is scanned and the L$_{\alpha1}$ decay process is recorded. These spectra benefit from a reduced spectral width with respect to total fluorescent yield data since a decay from an intermediate state with longer life time is detected. For RXES the incident energy is set resonantly to the energy of the Yb$^{2+}$ absorption line (here $h\nu$= 8938\,eV) and the energy transfer is scanned. This enhances strongly the Yb$^{2+}$ contribution and is therefore ideal for studying relative changes of the valence as function of an external parameter like temperature, doping or pressure.\cite{Dallera2002,Moreschini2007,Yamaoka2009,Rueff2011,Sato2014}  Experimental details are given in the \textsl{Appendix}.

In Fig.\,\ref{fig5}\,(a) the PFY-XAS spectra of Yb$_2$Si$_2$Al are shown for 25 and 300\,K. The data are analyzed by assigning identical lineshapes to the Yb$^{2+}$ and Yb$^{3+}$ contributions in the spectra, each with a tanh-type background and a background to account for the EXAFS above 8960\,eV. The PFY-XAS spectra clearly reflect the intermediate valent character of Yb$_2$Si$_2$Al. The valence at 300\,K obtained from the spectral weights in the PFY-XAS data differs slightly from the HAXPES analysis. We find 2.66(2) at 300\,K instead of 2.71(2) as in HAXPES. We ascribe the difference to the fact that the line shapes of the spectral weights in the L-edge are given by the empty 5$d$ density of states and are therefore less well defined.  At 25\,K the PFY data yield a valence of 2.61(2).

RXES data were collected for several temperatures between 25 and 300\,K. The spectra of some selected temperatures are shown in Fig.\,\ref{fig5}\,(b) and the insets show the regions of the Yb$^{2+}$ and Yb$^{3+}$ peak positions on a expanded scale. The decrease of the Yb$^{2+}$ and increase of Yb$^{3+}$ spectral weight with increasing temperature over the entire temperature range is clearly visible. The empirically assigned spectral weights give the valence $v_f$($T$) as function of $T$, however, due to the resonant enhancement of the Yb$^{2+}$ contribution these values have to be calibrated to the PFY-XAS values at 25 and 300\,K. The valence as function of temperature is displayed in Fig.\,\ref{fig5}\,(c). The valence changes from 2.61 at 25\,K to 2.66 at 300\,K, i.e. the overall change in the Yb valence is small and comparable with e.g. strongly intermediate valence Yb$_2$Ni$_{12}$(P, As)$_7$.\cite{Jiang2015}

\section{Discussion}

Our experiments are very clear that \YSA\ is an intermediate valence compound. The HAXPES data show that both Yb$^{2+}$ and Yb$^{3+}$ multiplets are present at room temperature.  The Yb valence of Yb$_2$Si$_2$Al is strongly non-integer, 2.68(2) at 80\,K, and 2.71 at 300\,K which points towards strong valence fluctuations even at room temperature. 
The Curie-Weiss temperature dependence of the magnetic susceptibility is relatively insensitive to the intermediate Yb valence, at least for temperatures larger than $\simeq$\,100\,K so that we identify the coherence temperature $T^*$ to be of the order of $\sim$100\,K, i.e. at larger temperatures, the Yb moments exhibit single ion behavior, below they approach a coherent state.

The low temperature Sommerfeld coefficient of the order of 0.1\,J/mol\,Yb$\cdot$K$^2$ in \YSA\ suggests a strongly correlated ground state. The combination of substantial valence fluctuations with strong correlations is a feature of Yb Kondo lattice compounds. This result may initially seem counterintuitive, since for  Ce ions the $f$-electron states are generally far from the Fermi level, and strong departures from integer valence are only possible when the hybridization is too strong to sustain Kondo physics. For Yb ions, the $f$-levels are in general much closer to the Fermi level, permitting substantial valence fluctuations even when the hybridization is weak enough to enable Kondo physics. The strongly enhanced Fermi liquid ground state and the substantial admixture of Yb$^{2+}$ states in the Yb ground state wave function indicate that this picture is appropriate for \YSA\, as it is for example in $\alpha$- and $\beta$-YbAlB$_4$,\cite{Okawa2010} YbFe$_2$Al$_{10}$,\cite{Khuntia2014} YbCuAl,\cite{Yamaoka2013} YbRh$_2$Si$_2$,\cite{Kummer2011} YbCu$_2$Si$_2$,\cite{Fernandez2012} or Yb$_2$Ni$_{12}$(P,As)$_7$.\cite{Jiang2015}

Further evidence for a
temperature scale that could be associated with Kondo compensation comes from the temperature dependence of the electronic specific heat $\Delta$$C/T$. As indicated in Fig.\,\ref{fig2}\,(a,c), there is a pronounced peak in $\Delta$$C/T$ that is centered at $T\simeq$\,32\,K that can be qualitatively described by the Bethe ansatz expression for the Coqblin-Schrieffer model,\cite{Rajan1983,Hewson1985} corresponding to a Kondo temperature in the range of 15\,K\,-\,32\,K, depending on the effective angular momenta of the Yb moments. The excess entropy $S^{\mathrm{mag}}$ associated with the onset of the strongly correlated electronic state and the peak can be determined by integrating $\Delta$$C/T$, i.e. $S^{\mathrm{mag}} = \int dT \Delta C/T$  (Fig.\,\ref{fig2}\,(c)). $S^{\mathrm{mag}}$ reaches the value R\,ln2 at a temperature near 60\,K, and using the Schotte criterio\,\cite{Desgranges1982} $S(T_{K})$\,=\,1/2\,R\,ln2, we estimate that the $T_{K}$\,$\simeq$\,32\,K, the same temperature where $\Delta$$C/T$ displays a maximum. 

The emerging picture is that with increasing temperature, the collapse of the strongly correlated state and the occupation of the excited state implied by the specific heat anomaly releases the entropy associated with a Kramers doublet.  The implication is that the Yb moments that lead to the Curie-Weiss susceptibility above 100\,K are effectively in a doublet ground state, where the other doublets that are created when the crystal field lifts the 8-fold degeneracy of the Yb$^{3+}$ ions are presumed to be at much higher energies.  With reduced temperature there are continuous modifications to the Yb valence, but there is sufficient Yb$^{3+}$ content that the Kondo effect remains energetically advantageous, even at temperatures that are as low as $\simeq$\,30\,K. 

\section{Conclusion}
Our objective is to identify a stoichiometric and low-dimensional  $R_{2}T_{2}X$ compound  where the exchange coupling $\cal{J}$ of the Yb-based $f$-electrons to the conduction electrons is relatively large. It has been previously deduced that $\cal{J}$ can be varied continuously in the series Yb$_{2}$(Pd$_{x}$Ni$_{1-x}$)$_{2}$Sn,~\cite{Kikuchi2009} and in Yb$_{2}$Pd$_{2}$(Sn$_{x}$In$_{1-x}$),~\cite{Bauer2004,Bauer2005,Bauer2010} where the Sommerfeld coefficient increases as the system passes from intermediate valence to Kondo, and then drops with the advent of antiferromagnetic order. While these systems provide the ability to tune the electronic scales continuously, the disorder that inevitably accompanies doping may make interpretation of the measurements problematic. In general, a stoichiometric compound is preferred. Like these compounds, the hierarchy of interion distances indicate that \YSA\ is also quasi-two dimensional, and in the absence of conduction electron hybridization could also be expected to display quantum mechanical excitations, particularly those associated with the collapse of the various exotic magnetic states that are predicted to occur in the SSL.  The experimental results presented here show that the Yb moments in \YSA\ are much more strongly coupled to the conduction electrons than those in Yb$_{2}$Pt$_{2}$Pb. The crucial question, as yet unanswered, is whether fractionalized excitations are also present in \YSA\, and if they are only slightly modified by coupling of the Yb ions to the conduction electrons, as in Yb$_2$Pt$_2$Pb, or if they are quenched, yielding conventional magnetic excitations familiar to us from heavy fermion and intermediate valence compounds.\cite{Fernandez-Alonso2015} Exciting new physics may be expected near the crossover between these two regimes, and future neutron scattering measurements will provide the means for placing \YSA\ within this range of possible behaviors.

\section*{Acknowledgment}
Work at Texas A\&M University (W. J. G. and M. C. A) was supported by NSF-DMR-1310008 and K.C., M.S., and A.S. benefitted from the financial support of the Deutsche Forschungsgemeinschaft (DFG) under grants SE-1441.

\section{Appendix}
\textbf{HAXPES:} The HAXPES experiments with $h\nu$=6.47~keV were performed at the Taiwan beamline BL12XU at SPring-8. The HAXPES spectra were taken by using a hemispherical analyzer (MB Scientific A-1 HE) and the overall energy resolution was set to 250\,meV for the Yb 3$d_{5/2}$, 3$d_{3/2}$ and valence-band spectra. Clean surfaces of the samples were obtained by cleaving \textsl{in-situ} under the base pressure of $10^{-9}$\,mbar at room temperature. The binding energy of the spectra was calibrated by the Fermi edge of a gold film.

\textbf{PFY-XAS and RXES:} PFY-XAS and RXES, as photon-in-photon-out techniques, provide probing depths of the order of 10 $\rm \mu m$, make them truly bulk-sensitive. PFY-XAS and RXES experiments were performed at the GALAXIES inelastic scattering end station at the SOLEIL synchrotron in France.\cite{Rueff2015} The synchrotron radiation was monochromatized using a Si(111) nitrogen-cooled fixed-exit double-crystal monochromator ($\Delta E/E\sim 1.4\times 10^{-4}$), followed by a Pd-coated spherical collimating mirror. The RIXS spectrometer was equipped with a Si(620) analyzer (R = 1m) operated around 76.7$^{\circ}$ Bragg angle.
Normalization was performed using a monitor placed just before the sample that recorded the scattered radiation from a Kapton foil placed in the beam path. The flight path was filled with helium gas to reduce the air scattering and the samples were cooled in a helium cryostat.

%\bibliography{Literature_Yb_2018_02_06_WJGedits}

\end{document}